# AN INJECTOR FOR THE CLIC TEST FACILITY (CTF3)

H. Braun, R. Pittin, L. Rinolfi, F. Zhou,  CERN, Geneva, Switzerland
B. Mouton, LAL, Orsay, France
R. Miller, D. Yeremian, SLAC, Stanford, CA94309, USA

*Abstract*

The CLIC Test Facility  (CTF3) is an intermediate step to demonstrate the technical feasibility of the key concepts of the new RF power source for CLIC. CTF3 will use electron beams with an energy range adjustable from 170 MeV (3.5 A) to 380 MeV (with low current). The injector is based on a thermionic gun followed by a classical bunching system embedded in a long solenoidal field. As an alternative, an RF photo-injector is also being studied. The beam dynamics studies on how to reach the stringent beam parameters at the exit of the injector are presented. Simulations performed with the EGUN code showed that a current of 7 A can be obtained with an emittance less than 10 mm.mrad at the gun exit. PARMELA results are presented and compared to the requested beam performance at the injector exit. Sub-Harmonic Bunchers (SHB) are foreseen, to switch the phase of the bunch trains by 180 degrees from even to odd RF buckets. Specific issues of the thermionic gun and of the SHB with fast phase switch are discussed.

## 1 INTRODUCTION

The CTF3 construction [1] will start in 2001. It is foreseen in 3 stages and the injector will also have 3 stages [2] of development. The first one is called "Preliminary stage". The injector is the same as the present one used for LIL [3] except for the thermionic gun. The second one is called "Initial stage". The injector has all new components (gun, bunching, focusing, and matching systems) and should deliver the nominal current with all RF buckets filled. The third one is called "Nominal stage". The injector is identical to the previous one except that SHBs will be added. Even and odd trains are generated with every other bucket filled. The concepts of fully-loaded linac and SHB with fast phase switch are explained in [1]. However the process produces unwanted satellite bunches.  The Preliminary stage is being implemented [4], so this paper focuses on the Nominal stage.

## 2 BEAM PARAMETERS

### 2.1 Target parameters at injector exit

The Preliminary stage is based on the existing LIL injector and therefore the beam parameters are based on experimental measurements.  For the Nominal stage, the charge in the satellite bunches should be as small as possible in order to get maximum RF generation efficiency and minimum radiation problems. The bunch length should be as short as possible to get both good beam stability (wake fields) in the following Drive Beam accelerator, and flexibility for the bunch compressor. The emittance should be minimised (linac acceptance, funnelling process in the combiner ring, decelerator linac). The single-bunch incoherent energy spread should be small in order to minimise bunch lengthening after bunch compression. The energy spread over the pulse, including beam-loading effects, should also be minimum. This will reduce the phase error after the bunch compression and therefore increase the 30 GHz generation efficiency. Table 1 summarises the beam parameters requested at the injector exit for the 3 stages.

Table 1: Target parameters at injector exit

| Parameters | Unit | Pre. | Init. | Nom. |
|---|---|---|---|---|
| Beam energy | MeV | 4 | $\geq 20$ | $\geq 20$ |
| Beam pulse | μs | 2.53 | 1.54 | 1.54 |
| RF pulse | μs | $\geq 3.8$ | $\geq 1.6$ | $\geq 1.6$ |
| Beam current | A | 0.3 | 3.5 | 3.5 |
| Gun current | A | 1 | 7 | 7 |
| Charge/bunch | nC | 0.1 | 1.17 | 2.33 |
| Bunch spacing | m | 0.1 | 0.1 | 0.2 |
| Bunches/pulse |  | 100 | 4200 | 2100 |
| Charge/pulse | nC | 10 | 4893 | 4893 |
| Charge/satellite | % | - | - | $\leq 5$ |
| Bunch length | ps- fwhh | 7 | $\leq 12$ | $\leq 12$ |
| Bunch length | mm-rms | 0.9 | $\leq 1.5$ | $\leq 1.5$ |
| Normalised rms emittances | mm.mrd | 50 | $\leq 100$ | $\leq 100$ |
| Energy spread (Single bunch) | MeV | $\leq 0.5$ | $\leq 0.5$ | $\leq 0.5$ |
| Energy spread (on flat-top) | MeV | - | $\leq 1$ | $\leq 1$ |
| Charge variation bunch-to-bunch | % | $\leq 20$ | $\leq 2$ | $\leq 2$ |
| Charge flatness (on flat-top) | % | - | $\leq 0.1$ | $\leq 0.1$ |
| Beam rep. rate | Hz | 50 | 5 | 5 |
| RF rep. rate | Hz | 100 | 30 | 30 |

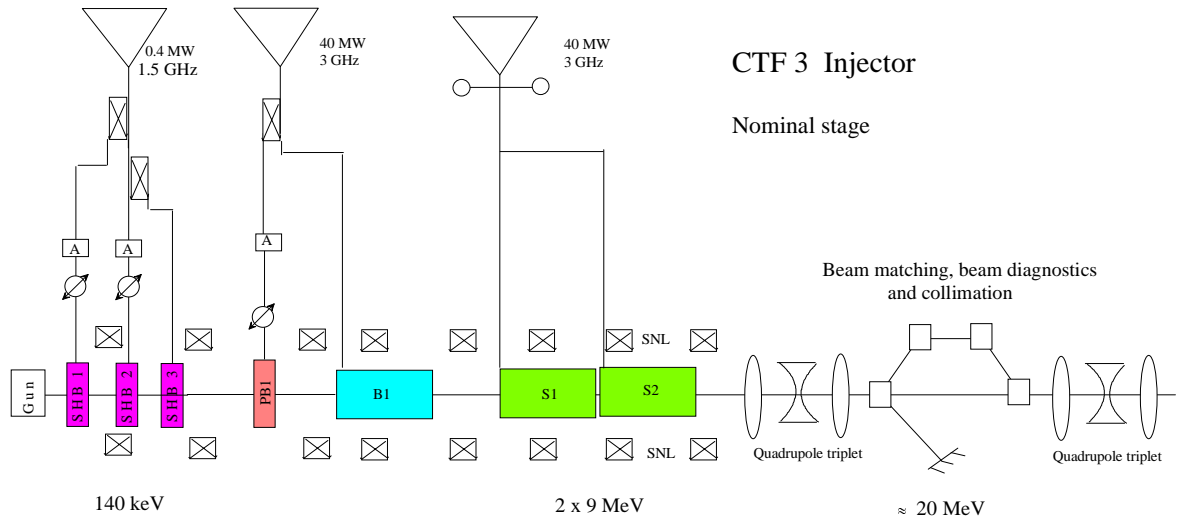

Figure 1: Layout of CTF3 injector.

## 2.2 Layout

Figure 1 shows the layout for the Nominal stage. After the thermionic gun (140 kV), there are 3 sub-harmonic bunchers SHB (1.5 GHz), one pre-buncher PB1 (3 GHz), one buncher B1 (3 GHz, 6-cells TW) and 2 accelerating structures S1/S2 (3 GHz, 32-cells TW). All components between the gun and the injector exit are embedded in a solenoidal field (SNL). A matching section with beam collimation and beam diagnostics is located between the injector and the Drive Beam accelerator.

## 3 EGUN SIMULATIONS

For the Preliminary stage, the new thermionic gun provides 2 A at 90 kV. A classical Pierce gridded gun, called CLIO [5], is proposed. It has a thermoelectronic dispenser cathode with an emitting surface of 0.5 cm$^2$, grid-cathode spacing of 0.15 mm, cathode-anode distance of 24 mm and anode hole diameter of 8 mm. The electrode geometry was modelled using the EGUN code [6]. The beam radius does not exceed 10 mm between the anode and 125 mm downstream of the anode, where a capacitive electrode allows beam current measurements. The normalised emittance is 7 mm.mrad, 62 mm downstream of the cathode. For the Initial and Nominal stages, the thermionic gun should provide 7A at 140 kV, with 150 kV voltage capability. A thermionic gun of the SLAC-type is proposed. It has a dispenser cathode with an emitting surface of 2 cm$^2$, cathode-anode distance of 45 mm and an electrode angle 45$^\circ$. Figure 2 shows the field lines and electron trajectories. Simulations in a space-charge-limited regime with thermal effect (1223 $^\circ$K) give a normalised emittance of 9.5 mm.mrad at z = 120 mm for a beam current of 7.4 A (perveance 0.128 μP). However, PARMELA simulations start at the anode exit. At this place, the normalised emittance is 7 mm.mrad. The maximum electric field on the contour is less than 10 MV/m.

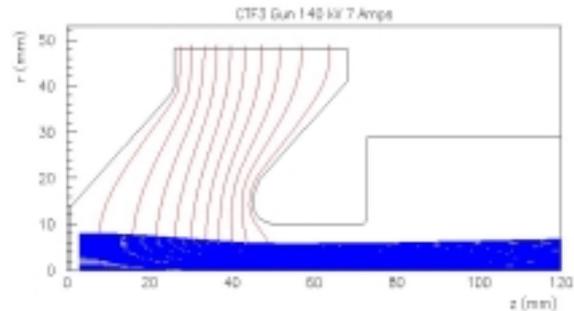

Figure 2: EGUN results for the CTF3 gun.

## 4 PARMELA SIMULATIONS

### 4.1 Longitudinal beam dynamics

Extensive simulations [7,8] are performed with PARMELA code at CERN, LAL and SLAC. They start from the gun exit with an initial normalised emittance of 7 mm.mrad, and 140 keV kinetic energy of the reference particle. The total number of particles is 6000 over a range of 6 S-band cycles. One of the issues is to make the satellite charge (in a 20$^\circ$ window) less than 5% of the main bunch (in a 20$^\circ$ window). Results are based on beam dynamics simulations assuming 400 kW RF power for 3 SHBs. Figure 3 shows the phase spaces obtained at the injector exit. The bunch length at the end of the injector is close to 10 ps (fwhh) and about 82% of the particles are captured in 20$^\circ$. Studies are going on based on a cluster of 3 sub-harmonic bunchers with 3 different frequencies: 1.5 GHz, 3 GHz, 4.5 GHz. Preliminary results, not discussed here, with a bunch length of 10 ps and a normalised rms emittance of 40 mm.mrad give 3% charge in the satellites.

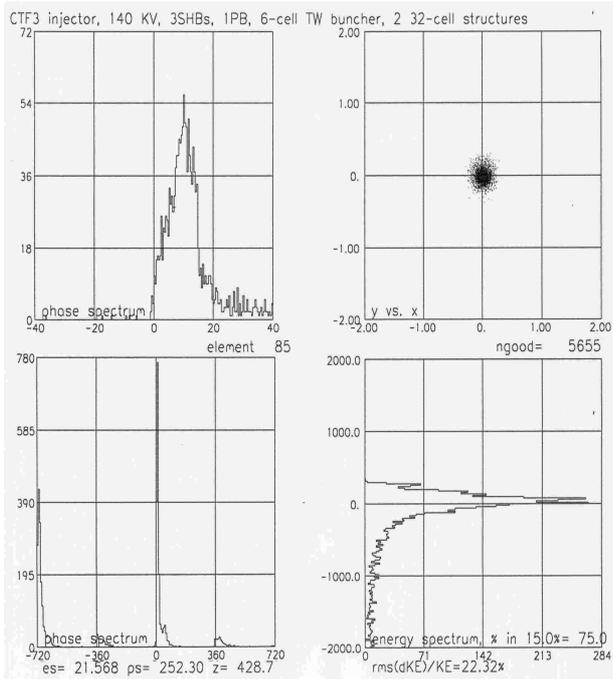

Figure 3: PARMELA results.
Left two figures: Particle number versus phase (degrees).
Right top: Beam size: y (cm) versus x (cm).
Right bottom: Energy spread (keV) versus particle number.

### 4.2 Transverse beam dynamics

To simulate the transverse beam dynamics correctly, different mesh sizes are chosen for different regions. Emittance variations are studied as a function of the strength and shape of the solenoidal magnetic field. An optimum is found for a value of 0.1 T along the 2 accelerating structures, with a maximum value of 0.2 T in the buncher. Table 2 compares the simulation results with the goal.

Table 2: Comparison between simulations and requirements

| Parameters | Simul. | Goal |
|---|---|---|
| Energy (MeV) | 20 | ≥ 20 |
| Satellite charge (%) | 5 | ≤ 5 |
| Bunch length (fwhh, ps) | 10 | ≤ 12 |
| Bunch length (fw, ps) | 20 | --- |
| Energy spread (fwhh, MeV) | 0.25 | 0.5 |
| Charge/bunch (nC) | 2.51 | 2.33 |
| RF power for 3 SHBs (kW) | 400 | Minim. |
| Normalised rms emittance (mm.mrad) (B= 0.1 T) | 33 | ≤ 100 |

## 5 GUN AND SUB-HARMONIC ISSUES

For the Preliminary stage the CLIO gun will replace the present one in the LIL tunnel. For the following stages (150 kV, 7A) voltage and current stability of ≤ 0.1% are requested on the flat-top. To obtain such performances, high voltage capacitors will be installed on a modified SLAC-type gun. Under these conditions the stored energy will be 1 kJ and a gun protection system has to be designed.

Concerning the SHBs 1.5 GHz, the PARMELA optimisation gives a gradient of 0.4 MV/m over a gap of 4 cm. This will require a voltage of 16 kV. Based on HFSS simulation [9], a power of 124 kW is needed at the input of each SHB. For 3 SHBs a total power of 372 kW will be necessary. A fast phase switch of the order of 3 to 4 ns is envisaged for the SHBs. It could be difficult to build a SHB with very low Q (~10) and to provide a 400 kW power supply at 1.5 GHz with large bandwidth.

## 6 CONCLUSION

The Preliminary stage of the CTF3 injector is being implemented. For the Nominal stage, a configuration of the injector has been found which fulfills the requirements of CTF3. The Initial stage should be easy to implement since it is a simple version of the Nominal stage. However, several steps will still be needed in order to improve the beam performance at the injector exit before starting the design of the RF cavities.

## ACKNOWLEDGEMENTS

The authors would like to thank I. Syratchev and R. Corsini for many improvements in the design. Stimulating discussions with J. Le Duff, G. Bienvenu, J. Gao, Y. Thiery (LAL) and R. Ruth (SLAC) are also acknowledged.

## REFERENCES

[1] CLIC study team, "Proposal for future CLIC studies and a new CLIC Test Facility (CTF3)", CLIC Note 402, July 1999.
[2] L. Rinolfi (Ed.), "Proceedings of the fourth CTF3 collaboration meeting", CLIC Note 433, May 2000.
[3] A. Pisent, L. Rinolfi, "A new bunching system for the LEP Injector Linac (LIL)", CERN PS 90-58 (LP), July 1990.
[4] R. Corsini, A. Ferrari, J.P. Potier, L. Rinolfi, T. Risselada, P. Royer, "A low charge demonstration of electron pulse compression for the CLIC RF power source", this conference.
[5] JC Bourdon et al, "Commissioning the CLIO injection system", SERA 91-23, LAL/LURE, Orsay, January 1991.
[6] W. Hermannsfeldt, "EGUN an electron optics and gun design ", SLAC, Report 331, October 1988.
[7] F. Zhou, H. Braun, "Optimisation of the CTF3 injector with 2 SHBs and 3SHBs", CTF3 Note 2000-01, February 2000.
[8] Y. Thiery, J. Gao, J. LeDuff, "Design studies for a high current bunching system", SERA 2000-130, LAL, Orsay, May 2000.
[9] I. Syratchev, private communication.